\documentclass[journal,11pt, onecolumn]{IEEEtran}
\usepackage{amsmath,graphicx}
\usepackage{algorithmicx}
\usepackage{algpseudocode}
\usepackage{algorithm}
\algdef{SE}[DOWHILE]{Do}{doWhile}{\algorithmicdo}[1]{\algorithmicwhile\ #1}
\usepackage{amsmath,graphicx}
\usepackage{dsfont, cuted}
\usepackage{amsfonts}
\usepackage[utf8]{inputenc}
\usepackage[usenames, dvipsnames]{xcolor}
\usepackage{subfigure}
\usepackage{flushend}
\usepackage{pifont}
\usepackage{multirow}
\usepackage{multicol}
\usepackage{booktabs}
\usepackage{lipsum}
\usepackage{fullwidth}
\begin{document}
\title{A Simulation Study to Evaluate the Performance of the Cauchy Proximal Operator in Despeckling SAR Images of the Sea Surface}

\author{Oktay~Karakuş$^{\dagger}$ \thanks{This work was supported by the Engineering and Physical Sciences Research Council (EPSRC) under grant EP/R009260/1 (AssenSAR).} \qquad Igor~Rizaev$^{\dagger}$ \qquad Alin~Achim$^{\dagger}$}

\author{Oktay~Karakuş,
Igor~Rizaev, Alin~Achim,
        \thanks{This work was supported by the UK Engineering and Physical Sciences Research Council (EPSRC) under grant EP/R009260/1 (AssenSAR). }
        \thanks{Oktay Karakuş, Igor Rizaev and Alin Achim are with the Visual Information Laboratory, University of Bristol, Bristol BS1 5DD, U.K. (e-mail: o.karakus@bristol.ac.uk; i.g.rizaev@bristol.ac.uk; alin.achim@bristol.ac.uk)}
}

\maketitle

\begin{abstract}
The analysis of ocean surface is widely performed using synthetic aperture radar (SAR) imagery as it yields information for wide areas under challenging weather conditions, during day or night, etc. Speckle noise constitutes however the main reason for reduced performance in applications such as classification, ship detection, target tracking and so on. This paper presents an investigation into the despeckling of SAR images of the ocean that include ship wake structures, via sparse regularisation using the Cauchy proximal operator. We propose a closed form expression for calculating the proximal operator for the Cauchy prior, which makes it applicable in generic proximal splitting algorithms. In our experiments, we simulate SAR images of moving vessels and their wakes. The performance of the proposed method is evaluated in comparison to the $L_1$ and $TV$ norm regularisation functions. The results show a superior performance of the proposed method for all the utilised images generated.
\end{abstract}

\section{Introduction}
\label{sec:intro}
\IEEEPARstart{S}{ynthetic} aperture radar (SAR) images of the sea surface provide useful information in a number of applications, including meteorology, environmental monitoring, wave structure analysis, ship monitoring, or energy generation. A common and important problem hampering statistical inferences from SAR imagery is the presence of multiplicative speckle noise. 
This may lead to loss of crucial details in SAR images and can cause problems in the analysis of these images, e.g. in feature detection, segmentation or classification \cite{kuruoglu2004modeling,achim2006sar,karakucs2018generalized}.

Since the availability of real SAR images is limited, more attention is being paid to SAR simulation. Simulated SAR images are especially useful when evaluating despeckling algorithms, in as much as a speckle-free image can be constructed. In order to form a SAR image, two types of modelling requires consideration: (i) sea surface modelling, and (ii) modelling of reflected SAR signal from the waves. There are numerous sea spectra available that have been developed based on experimental measurements \cite{arnold2007bistatic} for the modelling of the sea surface. Here, we adopt Elfouhaily et al. spectrum \cite{elfouhaily1997unified}, whilst for modelling the ship wake, the Kelvin wake model is implemented on the basis of Michell theory \cite{zilman2014detectability}. The elevation model of the water surfaces is generally considered as a superposition of ship-generated waves and wind-generated waves. For the simulation of SAR images, a two-scale composite model \cite{alpers1981detectability, romeiser1997improved} is used, including the tilt and hydrodynamic modulations, and the velocity bunching \cite{zurk1996comparison}.

The Cauchy distribution employed in this work is a member of the $\alpha$-stable distribution family and known for its ability to model heavy-tailed data. As a prior, it has a sparsity-enforcing behaviour, similar to its generalised-Gaussian counterpart, the Laplace distribution \cite{mohammad2012bayesian} (i.e. the $L_1$ norm), and it has generally been utilised in despeckling studies by modelling sub-band transform coefficients \cite{b:achim03a, chen2008wavelet}.

In this paper, we propose a despeckling study on simulated SAR images of the sea surface including ship wake structures. The proposed methodology incorporates the Cauchy distribution as a regularisation function. Furthermore, we derive the Cauchy proximal operator, which makes Cauchy regularisation applicable in standard proximal splitting algorithms such as forward backward (FB). Simulated SAR images incorporate a moving vessel with two different moving directions, and the performance of our despeckling approach is then compared to methods that use regularisation functions such as $L_1$ and total variation ($TV$) norms under log-normal speckle noise of different number of looks.

The rest of the paper is organised as follows: we first present our methodology for creating simulated SAR images of the sea surface and ship wakes in Section \ref{sec:sim}. In Section \ref{sec:Cauchy}, we introduce the Cauchy proximal operator and the despeckling methodology. In Section \ref{sec:experiment}, we present the experimental analysis, followed by concluding remarks and future work directions in Section \ref{sec:conc}.

\section{Simulation of Ocean SAR Images}\label{sec:sim}
The irregular sea surface model $Z_{sea}(x,y,z,t)$ with summation of many independent harmonic waves is formulated as
\begin{align}
   \sum_i \sum_j A_{ij} \cos \left[ k_i(x\cos\theta_j+y\sin\theta_j)-\omega_it+r_{ij} \right]
\end{align}
where $k_i$ and $\omega_i$ are the wavenumber and wave circular frequency, respectively, whilst $A_{ij}$ is the amplitude
\begin{align}
    A_{ij} = \sqrt{2S(k_i)D(k_i, \theta_j)dk_id\theta_j}
\end{align}
where $S(k_i)$ is the omnidirectional wave spectrum, $D(k_i, \theta_j)$ refers to the angular spreading function, $dk_i$ and $d\theta_j$ represent the sampling intervals. The omnidirectional Elfouhaily et al.~\cite{elfouhaily1997unified} spectrum is expressed as $S(k) = k^{-3}[B_l + B_h]$, with the gravity or the long-wave part of the spectrum
\begin{align}
    B_l = \dfrac{1}{2} \alpha_p \dfrac{c_p}{c} L_{PM} J_p \exp\left[ -\dfrac{\Omega}{\sqrt{10}} \left( \sqrt{\dfrac{k}{k_p}} - 1\right) \right],
\end{align}
the capillary or the short-wave part of the spectrum
\begin{align}
    B_h = \dfrac{1}{2} \alpha_m \dfrac{c_m}{c} L_{PM} J_p \exp\left[ -\dfrac{1}{4} \left( \dfrac{k}{k_m} - 1\right)^2 \right],
\end{align}
and the spreading function takes the form
\begin{align}
    D(k, \theta) = \dfrac{1}{2\pi} \left[1 + \Delta(k)\cos(2\theta) \right].
\end{align}

For modelling the Kelvin wake elevation, the fluid velocity potential $\varphi_{ship}$ and the ship elevation model relationship are used
\begin{align}
    Z_{ship} = \dfrac{U_s}{g}\dfrac{\partial\varphi_{ship}}{\partial x}
\end{align}
where the approximated form of fluid velocity potential (with parameters described in \cite{zilman2014detectability}) is presented as
\begin{align}
   \varphi_{ship}(x,y,z) = -\dfrac{16BL}{\pi}U_sFr^6 Re \int_0^{\infty}C(\tau,x,z)e^{iy\tau}d\tau
\end{align}

The scattering of the SAR signal from the disturbed surface of waves is a complex process, which involves the scanning platform geometry parameters (e.g. microwave signal parameters) and physical properties of the surface. According to the two-scale composite model (TSM), and taking into account the tilt and hydrodynamic modulations, the mean normalised radar cross-section (NRCS) is described as \cite{zilman2014detectability,romeiser1997improved}:
\begin{align}
    \Bar{\sigma}(x,y) = 8\pi k_e^4 \cos^4\mu_iW(k_{Bx},k_{By})|T|^2\times \left[1 + 2Re \int M(\mathbf{k})F(\mathbf{k})e^{i\mathbf{k}x}d\mathbf{k} \right]
\end{align}
where $k_e$ is the radar wavenumber, $\mu_i$ represents the local incidence angle, $W(\cdot)$ refers to the energy density spectrum of the surface roughness with components $k_{Bx}$, $k_{By}$ \cite{zilman2014detectability}. $M(\cdot)$ represents the complex modulation transfer function (MTF), and $F(\cdot)$ is the Fourier transform of the sea surface model. The final intensity image is obtained by applying the correction for nonuniform displacements of model facets in the azimuthal direction via the velocity bunching (VB) mechanism \cite{zurk1996comparison}.

\section{The proposed despeckling method}\label{sec:Cauchy}
Let us have an observed SAR image $G$ with multiplicative speckle noise $V$
\begin{align}\label{equ:multip}
    G = FV,
\end{align}
where $F$ is the speckle-free SAR image. The multiplicative image formation model given in (\ref{equ:multip}) is often manipulated as an additive one by taking the logarithm of  both sides $\log(G) = \log(FV)$, which leads to
\begin{align}\label{equ:additive2}
g = f + v,
\end{align}
where $g$, $f$ and $v$ refer to the logarithms of $G$, $F$ and $V$, respectively. The discrete wavelet transform (DWT) is a linear operation, which can consequently be applied to (\ref{equ:additive2}) to avoid the undesired effects of filtering operations. At each resolution level and for all orientations, we get additive terms corresponding to noisy wavelet coefficients ($\Gamma_{(i)}$) that can be written as the sum of the transformations of the speckle-free signal ($\Phi_{(i)}$) and the noise components ($\nu_{(i)}$) as
\begin{align}\label{equ:wavelet}
    \Gamma_{(i)} = \Phi_{(i)} + \nu_{(i)}.
\end{align}

\begin{figure}[htbp]
\centering
\includegraphics[width=.799\linewidth]{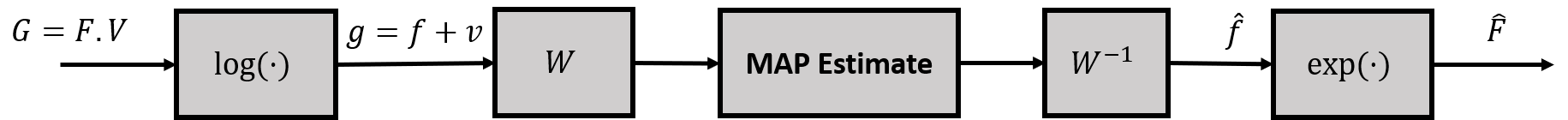}
\caption{Signal-dependent additive despeckling model in transform domain \cite{argenti2013tutorial}.}
\label{fig:despeckling}
\end{figure}

The despeckling model in this paper is depicted in Figure \ref{fig:despeckling}, where blocks $W$ and $W^{-1}$ represent the forward and inverse discrete wavelet transform operators.

In order to obtain the estimate of the speckle-free representation of the SAR image in the transform domain, $\hat{\Phi}_{(i)}$, the minimisation of the cost function with the Cauchy penalty term is then performed for each orientation $i = 1, 2, 3$ \cite{karakucs2020solving},
\begin{align}\label{equ:miniCauchy}
    \arg\min_{\Phi_{(i)}}\Bigg\{ \|\Gamma_{(i)} -  \Phi_{(i)}\|_2^2 - \sum_{j, k}\log\left(\frac{\gamma}{\gamma^2+\phi_{(i), j, k}^2}\right) \Bigg\}
\end{align}
where $\phi_{(i), j, k}$ refers to $\Phi(j, k)$ for orientation $i$.

In order to solve the minimisation problem in (\ref{equ:miniCauchy}) by using proximal splitting methods such as FB, \textit{the proximal operator} of the Cauchy regulariser should be defined. Basically, for any $\omega$-Lipchitz gradient function $h(\cdot)$ and $\omega>0$, the proximal operator is defined as \cite{combettes2011proximal}
\begin{align}
    prox_h^{\omega}(x) = \arg\min_u \left\{h(u) + \|u - x \|^2 / 2\omega \right\}.
\end{align}

Then, we substitute the function $h$ with the negative logarithm of the Cauchy distribution $-\log\left(\gamma/(\gamma^2+\Phi_{(i)}^2)\right)$, leading to the Cauchy proximal operator as \cite{karakucs2019cauchy1}
\begin{align}\label{equ:proxCauchy}
    \arg\min_u \left\{-\log\left(\frac{\gamma}{\gamma^2+u^2}\right) + \frac{\|u - x \|^2}{2\omega} \right\}
\end{align}

The solution to this minimisation problem can be obtained by taking the first derivative of (\ref{equ:proxCauchy}) w.r.t. $u$ and setting it to zero. Hence, we have
\begin{align}\label{equ:proxCauchy2}
    u^3-xu^2+(\gamma^2+2\omega)u-x\gamma^2 = 0.
\end{align}

Wan et al. \cite{wan2011segmentation} proposed a solution to the denoising problem of a Cauchy signal under Gaussian noise, and defined this solution as “Cauchy shrinkage”. Similarly, the minimisation problem in (\ref{equ:proxCauchy}) can be solved with the same approach as in \cite{wan2011segmentation},  however with different parameterisation. Hence, the solution to (\ref{equ:proxCauchy2}) can be obtained through Cardano’s method, which is given in Algorithm \ref{alg:proxCauchy}.

\begin{algorithm}[htbp!]
\caption{Procedure for Cauchy Proximal Operator}\label{alg:proxCauchy}
\begin{algorithmic}[1]
\Procedure{proxCauchy}{$x, \gamma, \omega$}
    \State $p \gets \gamma^2 + 2\omega - \frac{x^2}{3}$
    \State $q \gets x\gamma^2 + \frac{2x^3}{27} - \frac{x}{3}\left(\gamma^2 + 2\mu\right)$
    \State $s \gets \sqrt[3]{\frac{q}{2} + \sqrt{\frac{p^3}{27} + \frac{q^2}{4}}}$
    \State $t \gets \sqrt[3]{\frac{p}{2} - \sqrt{\frac{p^3}{27} + \frac{q^2}{4}}}$
    \State \textbf{return} $z \gets \frac{x}{3} + s + t$
\EndProcedure
\end{algorithmic}
\end{algorithm}

Having the Cauchy proximal operator, the minimisation for despeckling given in (\ref{equ:miniCauchy}), is subsequently solved via FB algorithm for any iteration $k$ as \cite{karakucs2019cauchy1}
\begin{align}
    u^{(k)} &= \Phi^{(k)}_{(i)} - \omega (\Phi^{(k)}_{(i)} - \Gamma_{(i)}),\\
    \Phi^{(k+1)}_{(i)} &= \textsc{proxCauchy}(u^{(k)}, \gamma, \omega).
\end{align}


\section{Experimental Results}
\label{sec:experiment}
\begin{figure*}[!ht]
\centering
\subfigure[Original image]{\includegraphics[width=.19\linewidth]{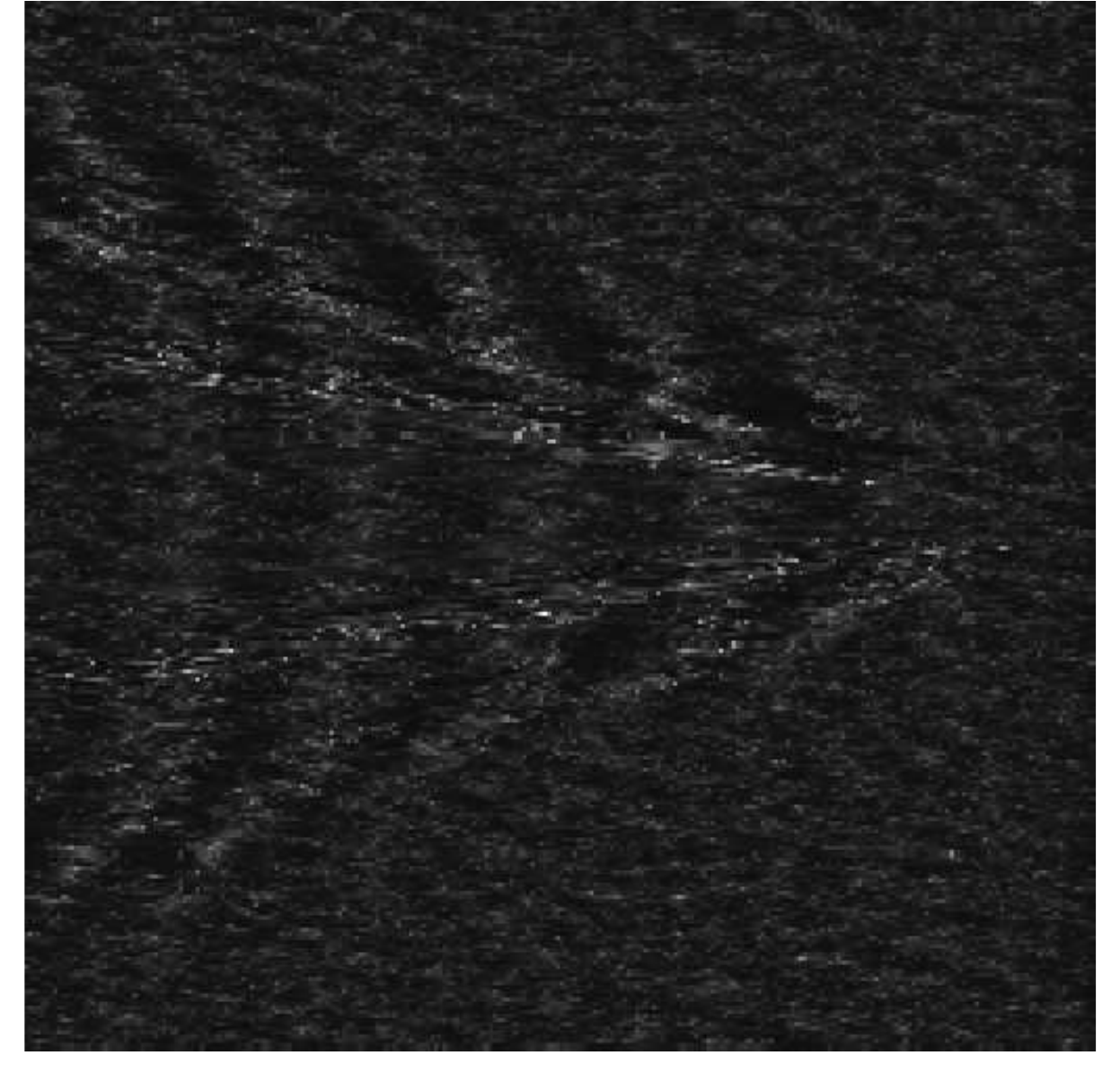}}
\subfigure[Speckled image (L=3)]{\includegraphics[width=.19\linewidth]{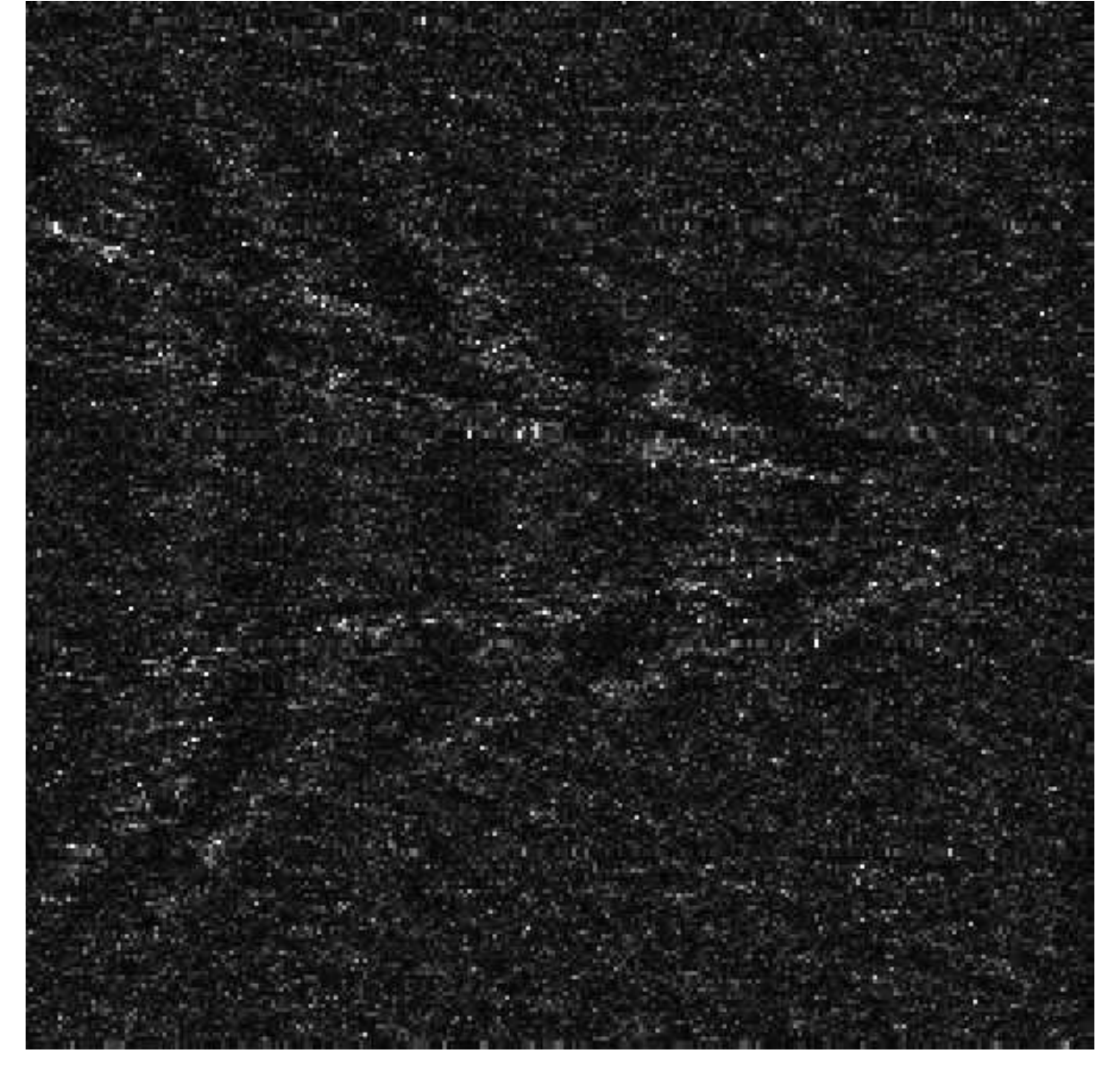}}
\subfigure[$L_1$]{\includegraphics[width=.19\linewidth]{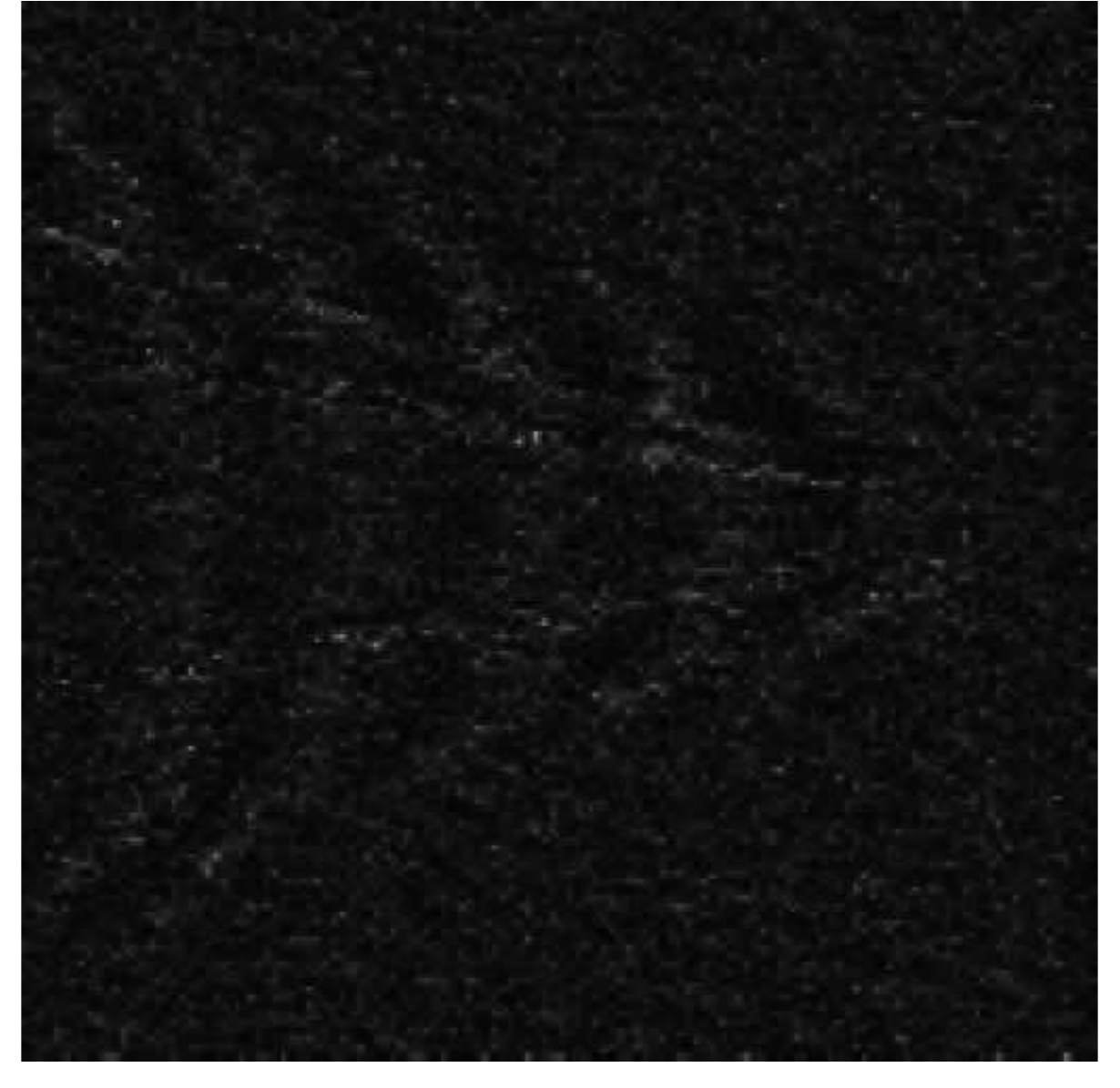}}
\subfigure[$TV$]{\includegraphics[width=.19\linewidth]{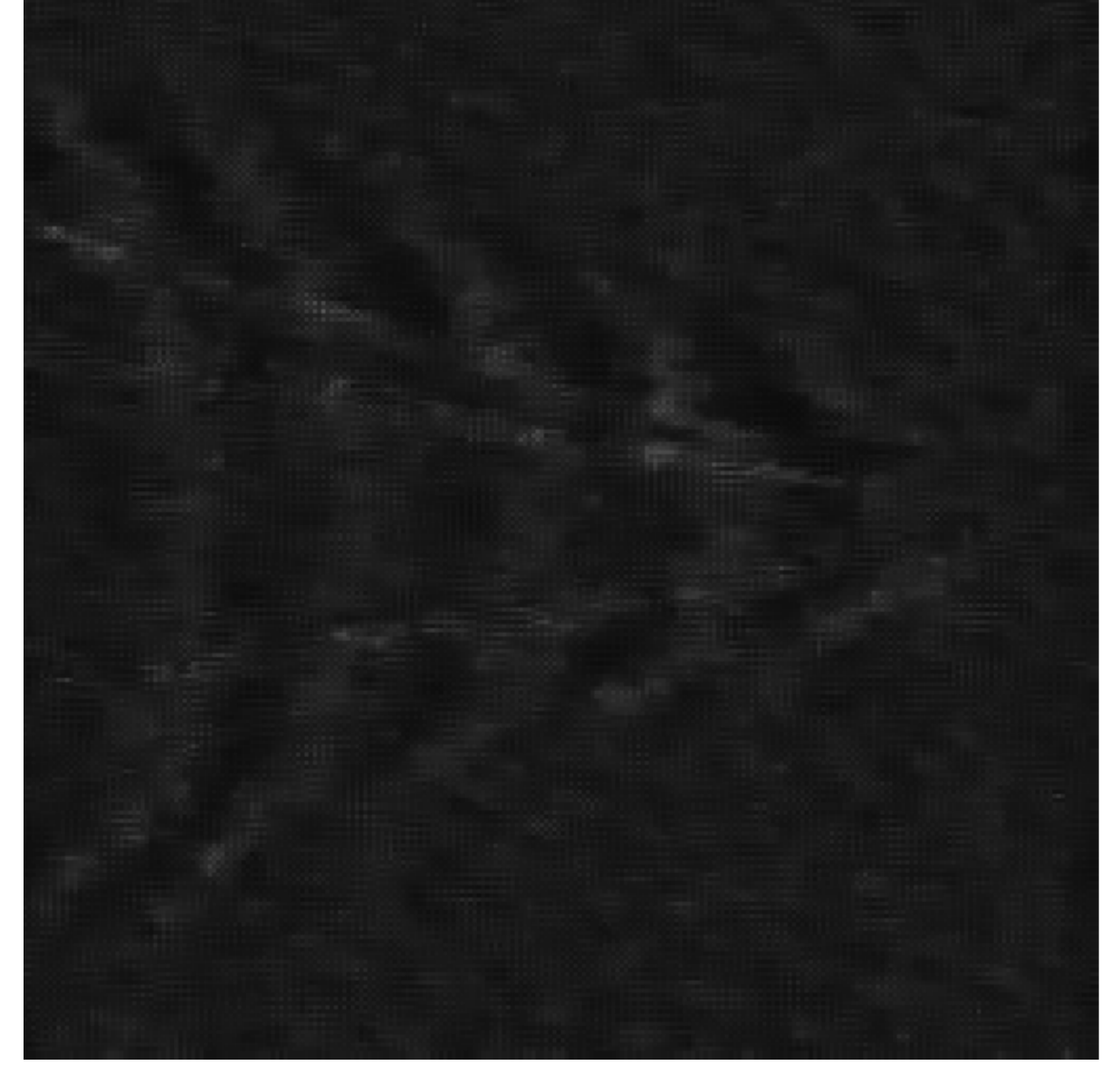}}
\subfigure[Cauchy]{\includegraphics[width=.19\linewidth]{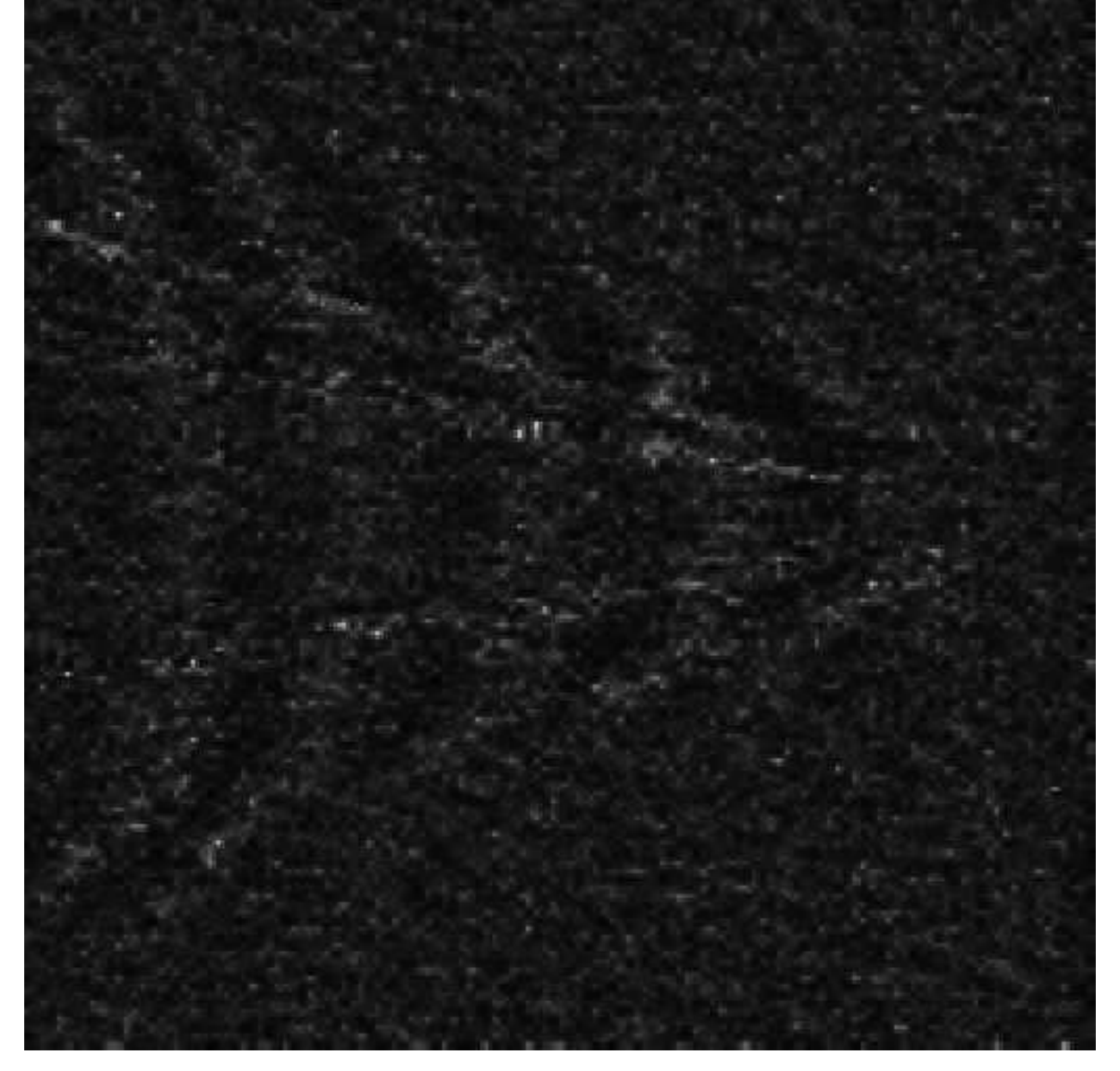}}\
\centering
\hspace*{3.5cm} \subfigure[Speckled image (L=7)]{\includegraphics[width=.19\linewidth]{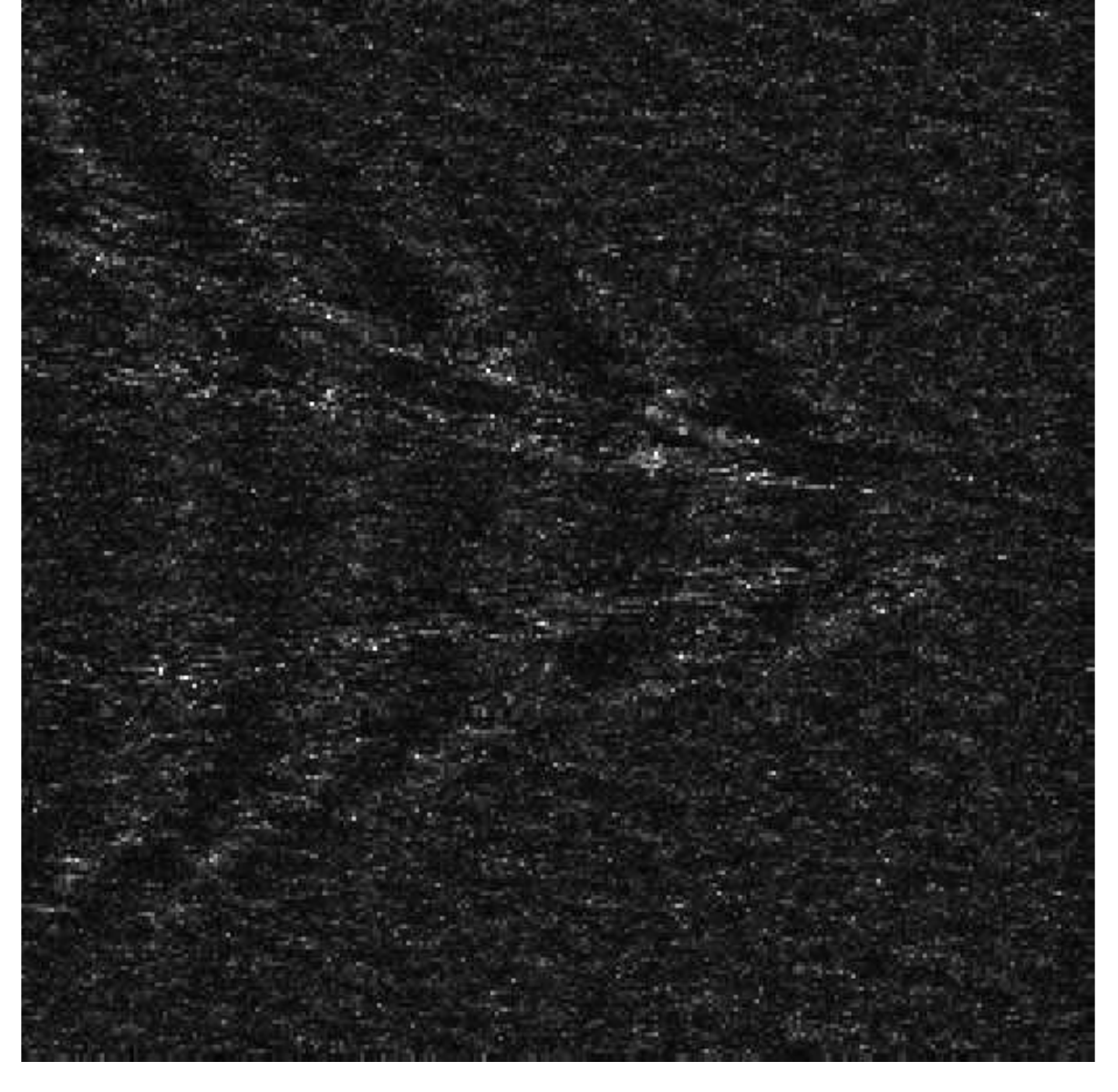}}
\subfigure[$L_1$]{\includegraphics[width=.19\linewidth]{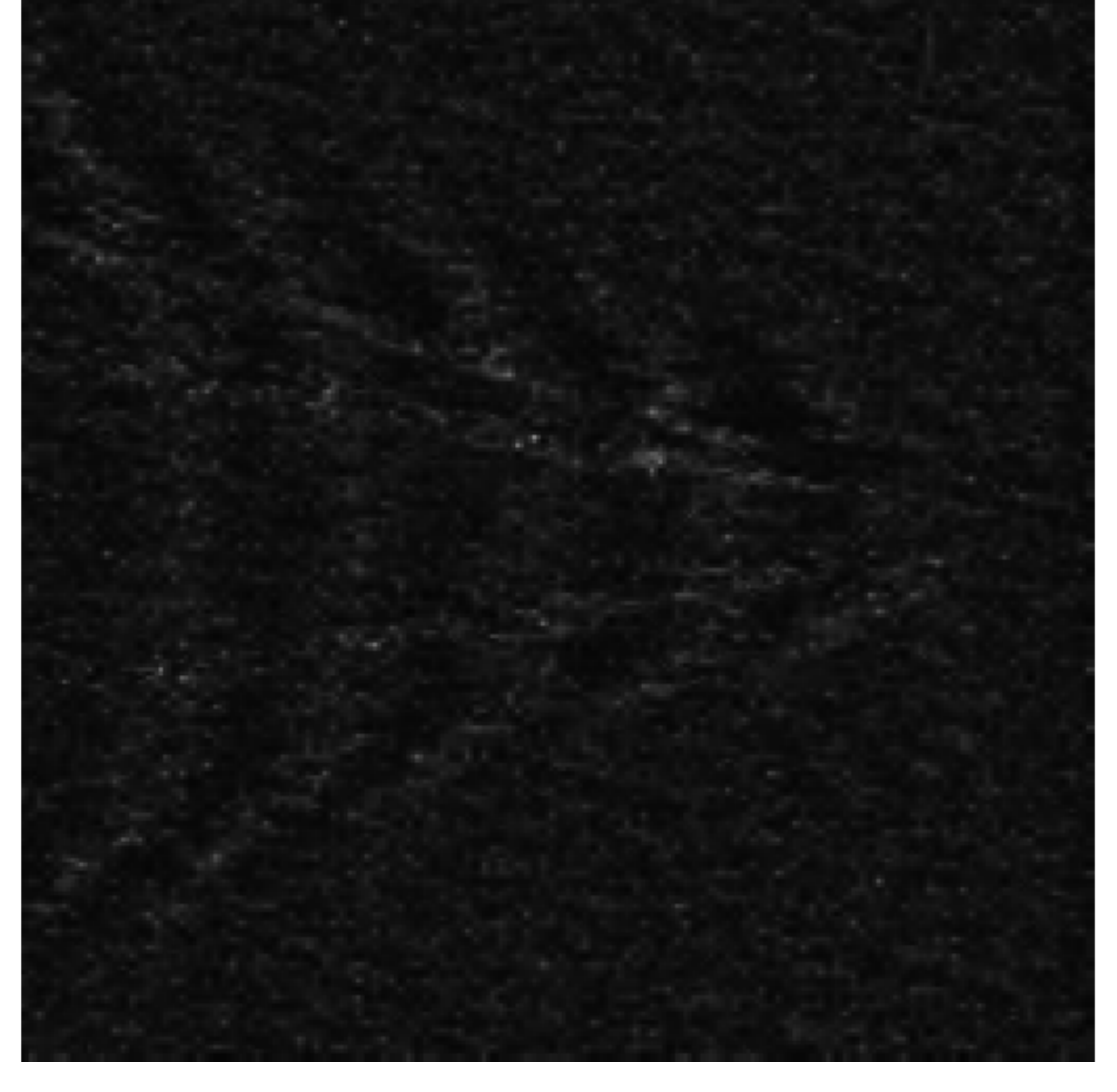}}
\subfigure[$TV$]{\includegraphics[width=.19\linewidth]{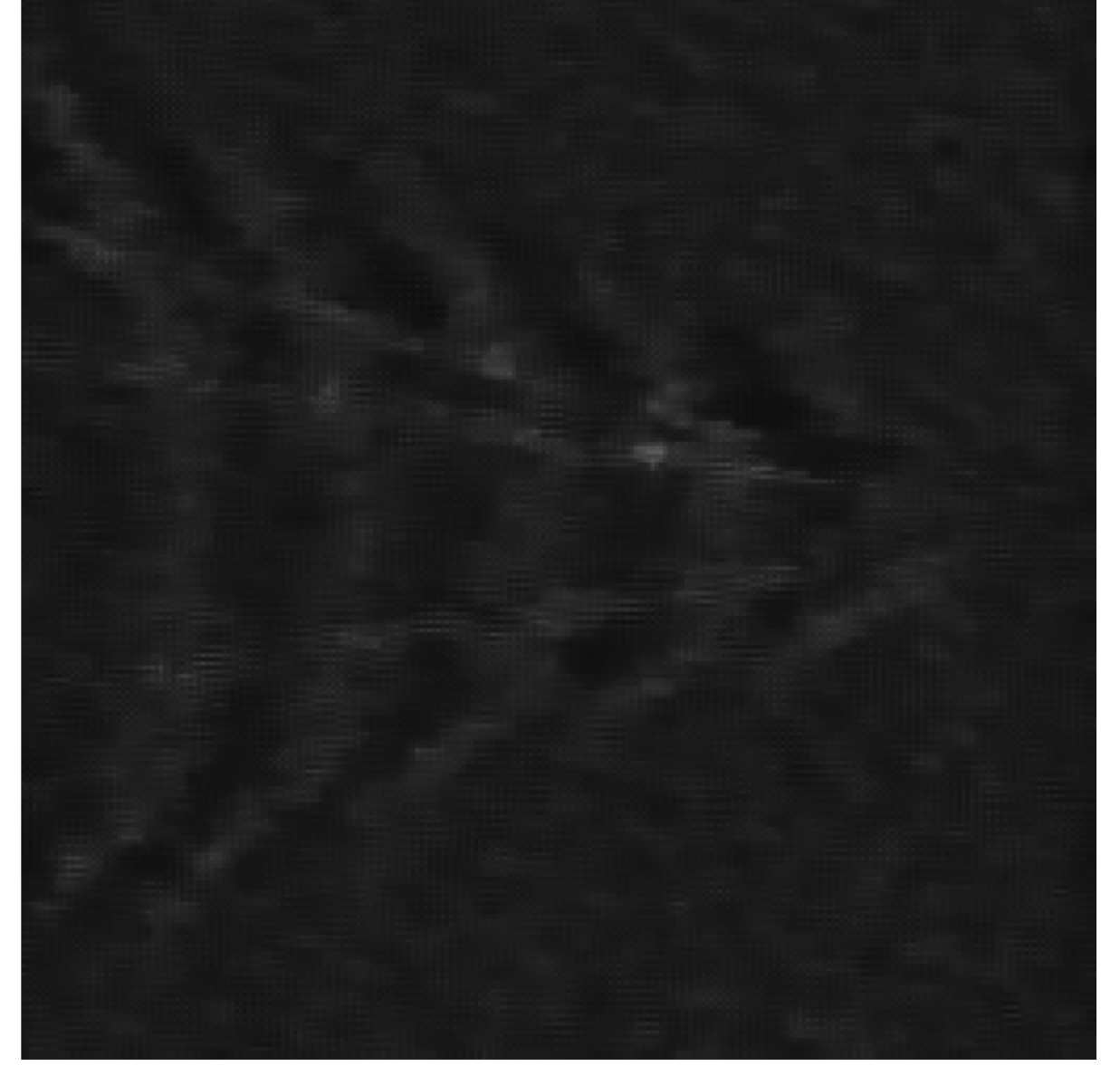}}
\subfigure[Cauchy]{\includegraphics[width=.19\linewidth]{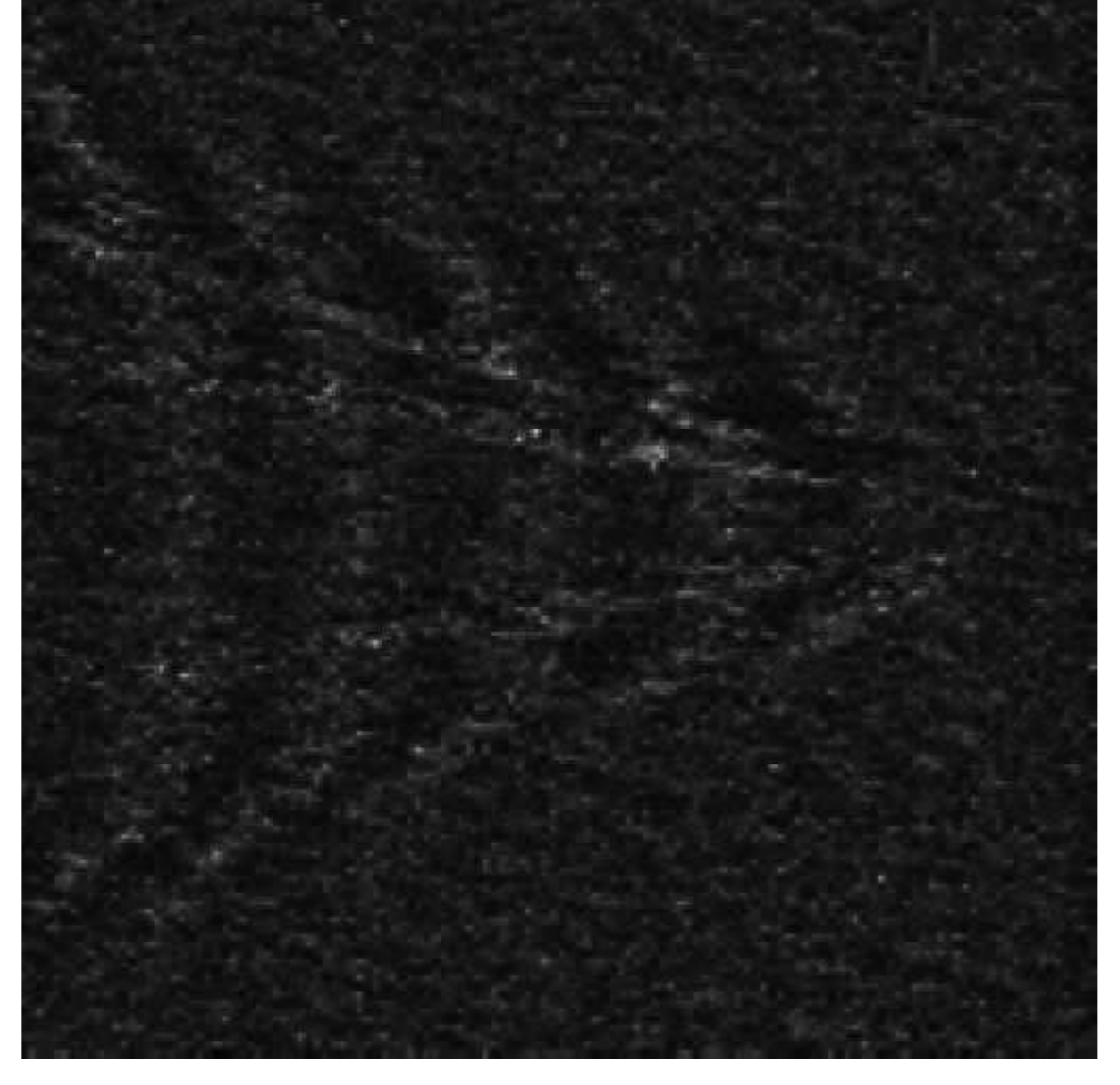}}
\vspace{-0.3cm}\caption{Visual despeckling results for Image-1.}
\label{fig:ds1}
\end{figure*}

The proposed method was tested for two different simulated SAR images, the main parameters of which are as follows. The size of scene is $512\times512$ m, the facet dimension for sea and ship wave models was set to 2 m. For both images, the wind speed was 5 m/s with direction of 45 degrees relative to the azimuth of the SAR platform. The parameters of the ship were selected as length $L = 52$ m, beam $B = 5.7$ m, and draft $D = 3.5$ m, and the Froude number, $Fr = 0.5$. The ship heading was 0 and 45 degrees relative to the azimuth for Image-1 and Image-2, respectively. The SAR platform parameters were: the platform altitude of 4.5 km, the platform velocity of 190 m/s, the SAR signal frequency of 9.65 GHz (X band) with VV polarisation, the incidence angle of 35 degrees, and lastly the azimuth and range image resolutions were both set to 2 m.

Both speckle-free simulated SAR images were then multiplied with the log-normal noise \cite{gagnon1997speckle} with the number of looks, $L$ chosen to be 3, 5 and 7. Speckle images for all three noise cases were processed by using the despeckling method for $L_1$, $TV$ and Cauchy regularisation functions. The performance of the methods were then compared in terms of peak signal-to-noise ratio (PSNR) and signal-to-mean squared error (S/MSE) values, which are given in Table \ref{tab:results}. In Figure \ref{fig:ds1}, despeckling results for Image-1 are depicted for $L$ values of 3 and 7.

Table \ref{tab:results} shows that the proposed Cauchy-based method achieved the best despeckling results for $L$ values of 5 and 7, whereas the $TV$ achieved better results for the case of $L = 3$. However, examining the visual results in Figure \ref{fig:ds1}, we can clearly see that despite its lower PSNR and S/MSE values for $L = 3$, the proposed method reconstructs both ship wake and sea surface structures and shows similar characteristics to the original speckle-free SAR image in Figure \ref{fig:ds1}-(a) when compared to $TV$. Even though $TV$ preserves ship wake structures as can be seen from the results in Figure \ref{fig:ds1}-(d) and \ref{fig:ds1}-(i), it discards the sea surface details whilst the final despeckled image is very blurry.

\begin{table}[ht!]
  \centering
  \caption{Despeckling performance measures}
     \resizebox{0.7999\linewidth}{!}{\begin{tabular}{rlrr|rr|rr}
     \toprule
          &       & \multicolumn{2}{c|}{L = 3} & \multicolumn{2}{c|}{L = 5} & \multicolumn{2}{c}{L = 7} \\
          &       & \multicolumn{1}{l}{PSNR} & \multicolumn{1}{l|}{S/MSE} & \multicolumn{1}{l}{PSNR} & \multicolumn{1}{l|}{S/MSE} & \multicolumn{1}{l}{PSNR} & \multicolumn{1}{l}{S/MSE} \\
          \toprule
   \multicolumn{1}{l}{Image-1} & Noisy & 22.363 & 4.785 & 24.484 & 6.995 & 25.920 & 8.474 \\
          & $L_1$    & 22.652 & 5.245 & 22.977 & 5.571 & 23.136 & 5.730 \\
          & $TV$    & \textbf{25.461} & \textbf{8.056} & 25.834 & 8.429 & 25.990 & 8.586 \\
          & Cauchy & 25.204 & 7.799 & \textbf{26.089} & \textbf{8.685} & \textbf{26.569} & \textbf{9.166} \\
          \midrule
    \multicolumn{1}{l}{Image-2} & Noisy & 25.126 & 4.792 & 27.274 & 6.987 & 28.712 & 8.460 \\
          & $L_1$    & 25.409 & 5.184 & 25.726 & 5.501 & 25.881 & 5.656 \\
          & $TV$    & \textbf{28.106} & \textbf{7.884} & 28.471 & 8.249 & 28.624 & 8.403 \\
          & Cauchy & 27.800 & 7.577 & \textbf{28.646} & \textbf{8.424} & \textbf{29.106} & \textbf{8.885} \\
          \bottomrule
    \end{tabular}}\vspace{-0.2cm}
  \label{tab:results}%
\end{table}%
\vspace{-0.5cm}

\section{Conclusion}
\label{sec:conc}
In this paper, we proposed a despeckling method, which we tested on simulated SAR imagery of the sea surface. Specifically, a closed form expression for calculating the proximal operator of Cauchy prior was proposed, which makes it applicable in proximal splitting algorithms such as FB. In addition, we presented a simulator for SAR images of the ocean surface that can incorporate ship wake structures. The performance of the Cauchy regularisation with FB algorithm showed better despeckling results both in terms of PSNR and S/MSE measures as well as visually for the simulated SAR images.


\bibliographystyle{IEEEtran}
\bibliography{refs}
\end{document}